# Rainbow: A Composable Coherence Protocol for Multi-Chip Servers

Lucia G. Menezo, Valentin Puente, and Jose A. Gregorio

**Abstract**—The use of multi-chip modules (MCM) and/or multi-socket boards is the most suitable approach to increase the computation density of servers while keep chip yield attained. This paper introduces a new coherence protocol suitable, in terms of complexity and scalability, for this class of systems. The proposal uses two complementary ideas: (1) A mechanism that dissociates complexity from performance by means of colored-token counting, (2) A construct that optimizes performance and cost by means of two functionally symmetrical modules working in the last level cache of each chip (D|F-LLC) and each memory controller (D|F-MEM). Each of these structures is divided into two parts: (2.1) The first one consists of a small loosely inclusive sparse directory where only the most actively shared data are tracked in the chip (D-LLC) from each memory controller (D-MEM) and, (2.2) The second is a d-left Counting Bloom Filter which stores approximate information about the blocks allocated, either inside the chip (F-LLC) or in the home memory controller (F-MEM). The coordinated work of both structures minimizes the coherence-related effects on the average memory latency perceived by the processor. Our proposal is able to improve on the performance of a HyperTransport-like coherence protocol by from 25%-to-60%.

——————————— ◆ ———————————

## 1. INTRODUCTION

Computing infrastructures are constituted by servers which typically employ multi-socket boards to increase computing density and utilization [1]. Additionally, in order to optimize yield, the cores in the device package might be distributed across Multi-chip Modules (MCM) [2][3][4][5]. Therefore, the available cores in the server are distributed across multiple chip multiprocessors/multicores throughout multiple packages in the server (from now on, *multi-CMP*). To optimize operational costs (by maximizing flexibility) and programmer productivity (by hiding complex underlying hardware details), a large proportion of the server market requires hardware support for shared-memory[6].

To reach this goal, hardware cache coherency is necessary. Nevertheless, maintaining both intra-CMP coherence and inter-CMP coherence is challenging [7]. Bandwidth availability for off-chip accesses is less than on-chip. In order to scale-up performance with the server sockets, off-chip resources should be scaled accordingly. This might increase the server cost. Similarly, to scale-up the count of cores, the size of the die should be increased, which might reduce the yield. Therefore, cost and performance seem hard to reconcile in this type of systems. Today scale-out seems to be the best choice to meet cost and performance requirements. Nevertheless, if we are able to design a system capable of circumventing inherent off-chip limitations, the trend might shift. As an example of this, recent commercial products [2] following a multi-CMP package approach have successfully increased the number of cores in the package at a fraction of the cost of the competition [8], producing affordable servers with 64 cache-coherent cores.

The main objective of our proposal is to find a balance between the three features required for this class of systems, namely: *complexity*, *scalability* and *performance*. Maintaining the coherence protocol complexity limited is an essential condition in order to achieve a feasible solution. However, this might inhibit certain performance optimizations. Similarly, multi-CMP physical limitations might prevent certain design choices for the coherence protocol.

For a given system size with enough interconnection bandwidth, relying on broadcasts enables good performance while keeping complexity of the coherence protocol constrained. For this reason, this has been a commonly used solution for many commercial systems [9][10][11]. Nevertheless, when the system's size grows or the production cost has to be constrained, directory-based coherence protocols are preferred [12]. Nevertheless, as well as the higher protocol complexity, the tracking mechanisms of these solutions have a cost overhead. These costs are amplified by the large, complex memory hierarchies of current server-oriented CMPs [13]. In this context, hybrid coherence protocols (broadcast-based and directory-based) are an interesting approach to avoid the drawbacks of both while retaining the desired properties [7][14][15].

The latency heterogeneity between the inter-chip and intra-chip communication reduces the appeal of a plain protocol across all the cores in the server. On the contrary, to embrace system latency and bandwidth heterogeneity, hierarchical coherence protocols seem more interesting [16]. However, the difficulty of their implementation is greater than in plain protocols. To couple two or more protocols into a hierarchy creates additional transient states and new protocol corner cases, which increase the verification complexity significantly [7]. Additionally, the overhead in storage to guarantee coherence invariance through all levels could be higher [17].

- L.G. Menezo is with the Department of Computing Engineering and Electronic, University of Cantabria, 39005 Santander, Spain. E-mail lucia.gregorio@unican.es.
- V. Puente Jr. is with the Department of Computing Engineering and Electronic, University of Cantabria, 39005 Santander, Spain. E-mail vpuente@unican.es.
- J.A. Gregorio is with the Department of Computing Engineering and Electronic, University of Cantabria, 39005 Santander, Spain. E-mail monaster@unican.es.

Although by exposing this to the programmer these problems could be somewhat alleviated, the coherence management problem is in part transferred to the system software and/or programmer sanity. Given the intricacy of the software stack of general purpose servers, which includes one or more levels of virtualization, containers, complex runtimes, etc. this design choice might be unattractive.

In this work, we propose a multi-CMP coherence protocol capable of dealing transparently with all the above-mentioned issues. First, token counting [18] is extended to obtain a composable system. It is used as a basis for isolating the complexity and the performance of the protocol. To read a data block, the processor will need one of the tokens associated with the requested block. Before the processor can write to a data block, all associated tokens must be collected. In contrast with Token-CMP [7], coherence is maintained separately between the intra-CMP and inter-CMP domains by adding three colors to the tokens (hence the name of the proposal).

To deal with cost and performance, the two structures, proposed originally in Flask Coherence [15] to minimize storage overhead, are distributed through the system coherency levels. One of these structures is basically a sparse directory [19] where only the most actively shared blocks will be tracked. The directory will be *loosely inclusive*, which means that if one of its entries is evicted, no external invalidation will be sent to the sharers. Since at a given time most of the data accessed by the cores is not being shared [20] even when the size of this directory is underprovisioned, the performance impact is limited. The second structure added is based on the use of a d-left Counting Bloom Filter (dlCBF) [21], which determines the presence (or absence) of a private block inside any of the caches of the system. If the filter is correctly dimensioned, the number of false positives (i.e. false presence of a copy of the requested block in any other chip) will be negligible. Therefore, the majority of the broadcast requests associated with the privately accessed blocks will be avoided. Only multicast requests for the first shared access of a block will be required (i.e. block level private-shared promotion) or if actively shared data are not being tracked by the directory (due to a previous eviction in it).

A complete implementation of our protocol proposal has been developed using a state-of-the-art simulation framework, Gem5 [22]. The implementation supports multiple 3-level cache CMP systems and it is capable of being used in a *full-system* simulation. The memory system is simulated using GEMS' memory simulation module Ruby [23]. As well as the current proposal, we have implemented a multi-CMP coherence protocol counterpart based on HyperTransport Assist. A complete implementation of both protocols is available in [24].

The results with a diverse set of workloads suggest that system performance may be mostly unaffected by the amount of storage devoted to maintaining coherency. In contrast, the counterpart protocol exhibits a high sensitivity to storage availability for such structures. Finally, these observations remain the same when the number of chips in the system is increased. These results indicate that the proposal might be an interesting approach to scale up system size with limited cost for the server or processor package.

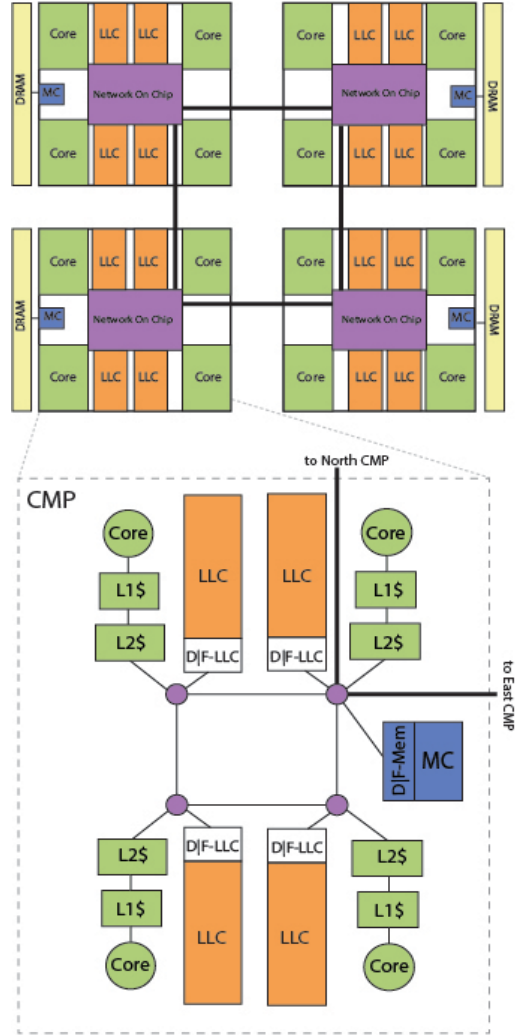

Fig. 1. Base architecture scheme. Example of four interconnected CMPs

## 2. MOTIVATION AND RELATED WORK

There is a plethora of coherence protocol proposals for multicore systems, each with its design goals (usually timeframe dependent). In the early days of CMPs, the solutions were based on using overprovisioned directories [4]. The large associativity required by these solutions limits their scalability. Not scaling this associativity in line with the private cache levels' associativity and/or the number of cores might induce artificial conflict misses (and subsequent external invalidations) in the directory. To reduce this cost some solutions make the sharer set encoding more flexible [25], removing the tag overhead [26], analyzing sharing patterns [27][28], using application semantics [14][20][29], or even hybridizing it with broadcast-based coherence approaches when the number of cores is limited [30][9]. In these broadcast-based approaches, to avoid both communication and tag snoop overheads, many authors advocate filtering [9][31], adapting the protocol behavior to the bandwidth availability [32][33] and/or providing on-network broadcast and/or gather support [34][35].

In contrast, the number of works focused on multi-CMP systems is more limited with a considerable diversity of the commercial solutions adopted.

The distributed caches in Haswell-EP-based systems [10] are kept coherent using a snooping-based protocol: in fact, two extensions of the MESIF protocol [36][37]. This protocol stores 2 bits of directory information to code 3 possible states for each cache line: remote-invalid, snoop-all and shared. This works like a filter to limit the snoop frequency. The protocol has two snooping modes: source snooping and home snooping. The snoop broadcast can be sent by the caching agent (a source snoop request), or by the home agent responding to receipt of the read request from the caching agent. The cores send requests to the caching agents in their node. In the case of an L3 miss, the caching agent forwards the request to the home agent, which provides the data from memory. Nevertheless, it can be sent in parallel and if the directory allows it, the home agent can forward from memory without waiting for snoop responses. Obviously, in this case snoop traffic is not reduced compared to the directory-less snoop.

AMD Systems employ Hyper-Transport Assist (probe filter) to avoid a large number of broadcast messages of broadcast protocols [9]. Up to 33% of L3 might be used as a sparse directory to track the sharers of the blocks which the memory controller is the home of. Although some policies try to avoid the eviction of shared lines, if the directory fills up, it has to evict an entry and invalidate any copy of the data block present in the private caches. Accesses to the DRAM and to the directory are done in parallel in order to reduce latency and the memory request is canceled if the entry is found in the directory. For some requests, such as a store to an S-state line in the requestor's cache, the request is always treated as a broadcast. In this protocol, the whole chip is considered as a node and the directory does not interfere in the internal transactions. Current AMD chips use a superset of Hyper-Transport called Infinity-Fabric, which seems to use an evolution of the same idea [38].

The IBM Power8 utilizes a non-blocking snooping protocol and snoopers respond in a fixed time employing time-division multiplexing [11]. However, its interconnect introduces the capability to oversubscribe each processor chip´s allotment of bandwidth. Systems built with Power8 can have up to 192 cores grouped in four groups of 4 chips each connected in an all-to-all manner. Each line has three scopes: chip, group and system. Commands issued with chip or group scope are not broadcast to the complete system, but rather only to the local chip or local group respectively. They utilize hardware prediction mechanisms to use the smallest scope to maintain coherence. If they mispredict, the request is retried with a different scope. This basically means that each data line includes MDIs (Memory-Domain Indicators). The system group is maintained in a coarse-grain directory called MCD (memory-coherence directory) included in the processor chip. Each MDI inside these MCD represents 16MB of granule memory. In any case, the coherence protocol is highly dependent on the characteristics of the processor and its interconnection network. The need for coherence maintenance has also extended to accelerators such as GPUs, FPGAs, and Smart Accelerators. Both proprietary solutions such as CAPI (Coherent Accelerator Processor Interface) for IBM Power-based systems [39], and more generic ones such as CCIX [40] that try to enable processors with different architectures to share data coherently with accelerators.

Analyzing the characteristics of different solutions, three essential aspects must be covered by any multi-CMP scalable solution. The first one is the necessity of separating, as far as possible, correctness from performance. Determining the correctness of a coherence protocol is a very complex task, so it must be isolated by a performance-enhancing optimization. A suitable method to achieve such task is token-based coherence. However, the basic solution shown in [18] and its extension to multi-CMP systems in [7], present a major problem with the starvation avoidance mechanism (*persistent requests*), which under adverse application conditions can ruin performance [41]. Our proposal uses tokens, but it eliminates the need for the above-mentioned starvation avoidance mechanism. Secondly, the protocol should reflect the different "areas" that constitute the multi CMP system. As was noted in [7], the protocol should not be a "flat" algorithm for the whole system, but nor should it have the complexity of a hierarchical one. The composability of our proposal by means of using colored tokens to circumvent the issue. Finally, as a third basic aspect to consider, it is clear that broadcast-based approaches do not scale with the number of cores, but neither do systems based on an inclusive directory which has to track/control the content of all caches in the system. It seems that an adequate solution can be found by using a hybridization of directory-based and broadcast-based protocols. For single chip multicores, Flask [15] exploited this idea successfully. Therefore, our proposal uses Flask as a starting point to tackle the multi-CMP systems coherence challenge.

## 3. RAINBOW COHERENCE PROTOCOL

### 3.1 Base System Architecture

The architecture used as a starting point is the one shown in Figure 1. It includes multiple CMPs which contain several cores, each with two exclusive private level caches, L1 and L2. A non-inclusive Last Level Cache (LLC) is shared by all the cores in the chip. In order to scale LLC bandwidth [42], this last level is divided into a set of banks interconnected with a Network-on-Chip (NoC). Associated with this level is a Directory/Filter structure (D|F-LLC) to control the coherence of the data blocks allocated in the private caches (it will be described in detail next).

Moreover, one memory controller (MC) is included for each chip. Associated with each MC, a second Directory/Filter structure is placed (D|F-MEM), which works in an analogous way to the D|F-LLC mentioned before, but containing information about the copies of the memory blocks owned by that controller throughout all system chips.

### 3.2 Support Structures for Rainbow

#### 3.2.1 Overview of D|F-LLC and D|F-MEM

In a coherent multi-CMP, a block might be private to one chip or shared among several chips. Simultaneously, inside each chip, a block can be private to one core or shared with other cores in the same CMP. This private-or-shared nature of the blocks is used to design the D|F-LLC and the D|F-MEM, by splitting each one of them into two different substructures.

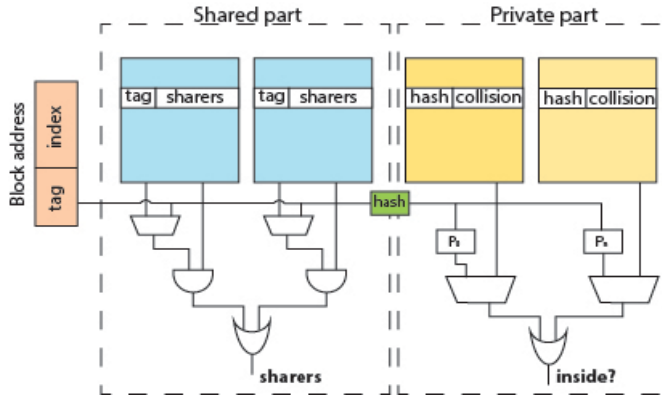

Fig. 2. Sketch of D|F-LLC and D|F-MEM structures.

The directory (D) will be dedicated to the localization of a portion of the shared blocks in the system. The filter (F) contains information about the blocks that are not being shared at all, i.e. private blocks. In both structures, these two modules work similarly although the information they hold concerns the cores inside the chip for D|F-LLC, but the chips in the system for D|F-MEM. Figure 2 shows a sketch of these structures

In the directory each entry includes the address tag and a bit for each of the sharers that have a copy of that block ("*sharers*" means cores in the D-LLC and chips in the D-MEM). Private blocks will not be tracked explicitly. Besides, these structures are not inclusive in relation to the cache content. This means that if any of the entries must be evicted from the directory due to a conflict, this can be done without having to invalidate all the copies which the entry is referencing, i.e. the data blocks present in the private caches.

When D-MEM/D-LLC has no information about a requested block, it might become necessary to broadcast a request to all potential coherence agents in the system. Both F-LLC and F-MEM are focused on reducing as much as possible unnecessary broadcasts. To do so, they will use a structure to store the information about the blocks that are present or not in any of the private caches in the F-LLC (or in any chip in the case of F-MEM). Using a probabilistic data structure, such as a *Counting Bloom Filter (CBF)*, will allow the storage requirements to be optimized, at the cost of having a non-null probability of false positives (by construction, false negatives in a CBF are impossible). A hit in F-LLC (after a miss in D-LLC) for a request, means that there is a copy in some private cache of the chip, and the request must be broadcast to all coherence controllers in the chip. A miss in the filter means that there is no other copy in the private caches inside the chip, and the request can be forwarded to the memory controller. Similarly, a miss in F-MEM (after a miss in D-MEM) means that there is no other copy of that block in any chip in the system and the request can be served from memory. A hit in F-MEM will require the broadcast of requests to all D|F-LLC and LLC controllers in the system.

When a block arrives for the first time at any private cache of the chip, the F-LLC's corresponding counter is incremented. When the block is evicted from the last private cache, the counter is decremented. Notice that if a block is only present in the LLC, the F-LLC will not contain its information. The LLC controller is aware that there is no other copy in the private caches via token counting, i.e. when a block has all the tokens. A similar structure is used to implement the F-MEM. Analogously its function is to determine the presence, or absence, of a block in any of the chips of the system. The counters are incremented when the block is requested by any chip and decremented when it is evicted from the LLC of each sharer.

In the D-LLC we will need space to store the tag and one bit for each of the cores inside the CMP. Any other type of shared representation [25][43] is compatible with this proposal. In the general case [12], to avoid external evictions it will be necessary to have as many entries as there are in the private caches. In our case, evictions are silent and consequently we can maintain the size of this directory constrained. Tracking the actively shared blocks is enough (i.e. kept in the private caches). Non-actively shared blocks can be fetched from LLC. D-MEM must track shared blocks that can be in either any private cache or in the LLC of all the different chips in the system. The size used for each filter in the F-LLC and the F-MEM will be set for a given probability of false positives with limited effect on the system performance [21][15].

F-LLC counter updates are done after private cache misses or private caches evictions; in both cases overlapped with LLC access. Similarly, F-MEM updates are done overlapped with on-memory accesses and cache evictions. On-chip clean block evictions should be notified to the home memory controller. For implementing F-LLC and F-MEM, we will use a d-left Counting Bloom Filter (dlCBF), introduced by Bonomi et al. [6], which uses modified Counting Bloom Filters (CBF) that at least double the efficiency of regular CBF counters with a similar implementation cost [15].

### 3.3 Composable Token Counting

In order to maintain the Single-Writer Multiple-Reader (SWMR) invariant, *Rainbow* uses token counting. However, to hide multi-CMP's complex nature, we have included an extension to the flat policy that comes by using *colored* tokens. This way, associated with each data block, there are three types of tokens: (a) one **gold** token; (b) as many **silver** tokens as the number of chips in the system; (c) as many **bronze** tokens as the number of cores in the system. *Bronze* tokens are functionally equivalent to the original tokens of Martin et al [18], i.e. only one is needed to read the data block but all of them are required to write in that block. The other two colors can be seen as an extension of the '*owner*' token in flat token protocols, that is, to be the only one replying to some of the requests under some specific circumstances, which will be explained next. Like the Infinity Fabric coherence substrate, the token coloring scheme is applicable with chip heterogeneity (both in core count and memory controllers per chip). Here we assume a homogenous system, to facilitate the discussion and comparison with other alternatives.

Initially, the home has the block with all the tokens. When a core from CMP1 issues a request to *read*, the memory controller sends the data block requested with all its tokens (*gold*, *silver* and *bronze*). From that moment, the requestor cache controller will have the *golden* token so it will be in charge of replying to external read requests for that address. Since it is also holding at least one *silver* token, it will also be in charge of replying to the read requests from the cores inside the same chip (via the corresponding cache controller). When a read request from another CMP2 arrives, the *golden* holder will send one of its *silver* tokens and as many *bronze* ones as cores the requestor CMP2 has (which can be different to the cores in CMP1). This will make the new requestor the *silver* owner of its chip and so the one in charge of replying to further requests from the cores in it. Note that the *golden* token holder is still CMP1. When the request is a ***store***, all the tokens distributed through the system must be collected by the requestor coherence controller. Next, taking advantage of the D-MEM/D-LLC structures described in the previous subsections, we will explain how this process is done.

### 3.4 Protocol Actions

#### 3.4.1 Read miss

After a miss in the private cache levels, the request is forwarded to the D|F-LLC and LLC controller simultaneously (Figure 3). If there is a hit in the D-LLC, that block is/has been accessed by multiple cores. In this case, a request is sent to the *silver* token owner (which must be annotated in the sharers' vector) and the new requestor is added to the sharer list. If there is a miss in the D-LLC, and a hit in LLC (with at least one bronze token), the block is sent to the requestor. If there is also a miss in LLC, a lookup is issued to the F-LLC filter. If there is miss in the filter, there is guarantee that there is no copy in the chip. Consequently, the request will be forwarded to the corresponding memory controller. A hit in F-LLC potentially indicates that there is at least one copy of the block allocated in other private caches of the CMP. In this case, three situations are possible: *(1)* it might be a shared block previously evicted from the directory; *(2)* it is a private to share promotion, i.e. the block has previously been accessed by another core in the CMP; and *(3)* it is a false positive in the filter.

After a filter hit, the controller will issue a multicast to all the private cache controllers inside the chip requesting the *silver* token holder to send a copy of the data block along with at least one *bronze* token. The private controllers will reply to the LLC coherency controller with the number of tokens that they have for that block in their private caches (including zero). Subsequently, after silently evicting one entry, a new entry for that block will be allocated in the D-LLC. If none of the cores has any token (i.e. a false positive happened), then the D|F-LLC will forward the request to the home memory controller, which will start a similar process in D|F-MEM structures. Therefore, the potential performance impact of this approach is that true memory accesses will be slightly delayed (until the sharing information is reconstructed). With properly sized filters this probability is fairly small [6].

#### 3.4.2 Write miss

To guarantee single writer invariant, the requestor needs to collect all the tokens (*golden*, *silver* and *bronze*) assigned to that data block before reaching the exclusive state. Similarly, the

```
if ( shared_data ) then
    //entry in D-LLC
    request to the silver-token owner;
else if ( data_in_LLC ) then
    send data+tokens to the requestor;
else if ( block present in chip ) then
    //present in F-LLC
    broadcast;
    creation of new entry in D-LLC;
else
    // data out of the chip
    read-request to the D-MEM home;
```
Fig. 3. Read miss

```
if ( shared_data) then
    //entry in D-LLC
    multicast to chip-sharers to invalidate;
    if ( all_tokens_are_present ) then
        complete the request;
    else
        send request to D|F-MEM home;
else if (data_in_LLC_with all tokens ) then
    send data to requestor
    update F-LLC
else if ( block present in F-LLC ) then
    broadcast;
    creation of new entry in D-LLC;
    complete request;
else
    //data out of the chip
    write-request to the D-MEM home;
```
Fig. 4. Write miss

```
if ( no_data_copies_outhere ) then
    read_memory;
else
    if ( data_shared_by_chips ) then
        multicast to chip sharers
    else
        //data present in chips, but no info where
        broadcast to all chips;
        create new D-MEM entry;
```
Fig. 5. Actions for requests to D-MEM

request is sent to the D-LLC and LLC controller. There are four different outcomes according to how the tokens are distributed (Figure 4). If the D-LLC has an entry indicating that the block is being shared inside the chip, it requests all the tokens from the sharers (i.e. invalidating their copies). If the D-LLC receives the data block and all its tokens, they are all forwarded to the requestor and the directory entry is updated with the new information. If another chip in the system holds a copy of the block, the controller will be aware simply by realizing that some tokens are missing. In this case, the request is forwarded to the corresponding D|F-MEM in order to collect the missing tokens.

If there is no entry allocated in the D-LLC, but the LLC has the data block with all the tokens, then these are forwarded to the requestor. In this situation, the corresponding counter in F-LLC must be updated for that address, indicating that the block is inside a private cache. No entry is allocated in the D-LLC for this block since at the time of the request it is not being shared.

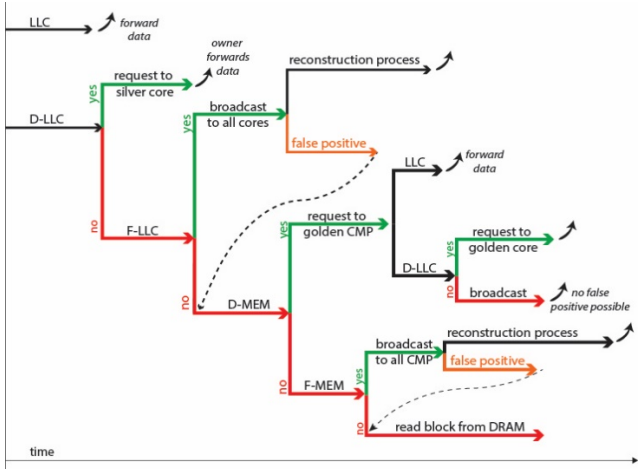

Fig. 6. Timeline of a read miss and potential outcomes.

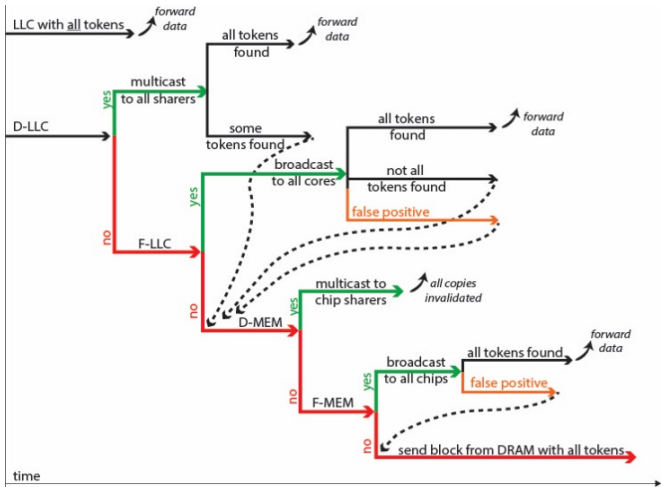

Fig. 7. Timeline of a write miss and potential outcomes. (For readability reasons, the time-scale is not proportional)

Similarly to the reads, if there is a miss in D-LLC and in LLC, but a positive lookup in F-LLC, then a broadcast is sent to all the cores in the chip (this time requesting all tokens, not only information about them). If not, all the tokens are collected and the request is also sent to the D|F-MEM home controller, which will start the token collection process for all or part of the CMP in the system.

When a request arrives at the D|F-MEM, the process is similar (Figure 5): if there is a hit in D-MEM, the CMPs included in the sharers' list will be notified to forward their tokens and the process will be repeated in each D|F-LLC. If there is a miss in the D-MEM and a hit in the F-MEM, the request will be forwarded to all CMPs in the system. If there is a negative in F-MEM, it means that the processor is trying to access a block with no copies in the system and so it can be forwarded directly with all tokens. The request leaves the chip only when it must. In turn, when D|F-MEM requests are sent to all the chips, they will use their coherence information at the D|F-LLC to decide what to do with that request (unicast, multicast or broadcast). External requests (i.e. from memory controller) are handled like on-chip access, therefore the probability of unnecessarily snooping the private caches in the system is fairly low (we need a combination of false positives across all filters).

Colored token counting plays a key role in this protocol favoring its composable nature. If the block is present inside the chip, there is a local *deliverer* (the one holding the silver token) which will avoid read requests going outside the chip. In the write requests, token counting allows the requestor to stop waiting for the answers when all the tokens have arrived. This avoids on-chip broadcasts when the data block is present in the LLC with all the tokens, indicating that it is the only copy of the block in the CMP or in the system. Note that this is a very usual case since most of the blocks will be evicted from the private caches to LLC. Unnecessary off-chip traffic is also saved when detecting that all the tokens are present inside the requesting chip, which ensures there are no more copies in other chips of the server. Additionally, token counting facilitates to infer when to unset an address in the corresponding F-LLC or F-MEM.

### 3.4.3 Updates and replacements

When a block is sent from the corresponding memory controller for the first time, the F-MEM information for that address is updated to indicate that a copy of the block is present in a chip of the server. D-MEM is still not modified, as the block is assumed to be "chip-private", i.e. it is held by one chip only (the one requesting it). The memory controller is oblivious to any other core accessing the data in the same chip. When that block is requested by another chip, a new entry must be allocated in the D-MEM since the block is then considered shared. The controller follows a similar approach when a block is promoted from private to shared, i.e. a multicast will be issued to all LLC controllers after a F-MEM hit.

The new entry will silently evict the tracking information of the corresponding previous address. Therefore, although the sharing information is lost, the sharers will not have to invalidate their copy of the block. Subsequent shared accesses to that block will 'reconstruct' the sharing information of D-MEM on-demand.

Finally, when a block is replaced from all the private caches, the coherence controller removes the corresponding information in the D-LLC: sharing information (if present), and the counter associated with the address decremented in the F-LLC corresponding counter. To evict a block from LLC we need to collect all the tokens in the chip. If a *bronze* token is missing, this means a private cache holds a copy (which might also have *silver* or *golden* tokens). The most likely scenario is that at eviction time LLC entry will hold all the tokens in the chip. In practice, capacity ratios between LLC and private caches make this case unlikely. Therefore, the solution adopted is to invalidate all copies of the block in the chip. The block eviction is handled by D|F-MEM: if the block has all the *silver* and *golden* tokens, the information will be sent to the memory controller, updating the filter counters and directory accordingly. If some silver or golden tokens are missing in the incoming block, a system wide invalidation is sent for the block address (handled like write operations).

### 3.4.4 Protocol timeline

Next, a summary of the main actions of the protocol for read and write misses in private caches will be provided. Since some actions take place in parallel, while others are sequential, a temporal representation has been provided. Figure 6 and Figure 7 show the timeline of the events and actions behind a read and a write miss in private caches, respectively. The representation corresponds to an excerpt from the full specification of the protocol. We present the most frequent scenarios after a miss in private cache, including requests and responses on and off chip. Due to space limitations, the full protocol specification (including corner cases not depicted here) is available to the interested reader in [24].

## 4. EVALUATION METHODOLOGY

### 4.1 System Configuration

To evaluate our proposal, we use different multi-CMP configurations with two and four chips. The 4-chip configuration is like the one shown in Figure 1. Each chip includes 4 cores interconnected with a mesh topology. A summary of the main parameters of all the elements can be seen in Table 1. Some characteristics will be modified to determine their impact on the performance of the proposed protocol.

There are three levels of cache. The first two levels are private, exclusive (i.e. L2 acts as a victim cache of L1). The third level is shared by all the cores of the chip and uses a mesh network. This level (and the directory of D-LLC) is banked and interleaved by the least significant bits.

We will analyze the sensitivity of the most significant parameters in the performance and scalability of our protocol.

### 4.2 Simulation stack and Workloads

In order to perform full-system simulations, we use gem5 [22] as the main tool for our evaluation. Although the cost of this decision is substantial, it increases confidence about the feasibility of the proposal. The simulated system is based on a x86-64 platform, runs Debian8.0 with a Linux 4.1 kernel, and Docker containerized workloads. KVM assisted simulation has been used to fast-forward the state of the system to the region of interest for each workload [44].

Our proposal, as well as the counterpart protocol, have been implemented using SLICC language (Specification Language for Implementing Cache Coherence) which is part of the GEMS memory simulator [23], currently integrated in the Gem5 infrastructure [22].

We choose a representative set of benchmarks, enumerated in Table 2. Some are multi-programmed workloads from the SPEC CPU2006 benchmark [45] running in rate mode with reference input sizes. The numerical applications are from the NAS Parallel Benchmark suite (OpenMP implementation [46]). Multi-threaded workloads have been chosen from the Princeton Application Repository for Shared-Memory Computers (PARSEC v3.0) [47]. The mix of workloads has been selected trying to cover diverse usage scenarios, varying the sharing degree and the sharing contention.

### 4.3 Counterpart: Hyper-Transport like coherence protocol

To compare our proposal, we have implemented a coherence protocol based on the one present in the AMD Opteron [9]. This includes a broadcast-based coherence protocol where the last-level cache misses are redirected to the memory controller home. This home node has a cache directory (probe filter) which includes information about all cached data in the system (from the cache lines in its memory). When a request is received, the memory controller issues a broadcast, a single request or no request at all, depending on the information it has in the directory. This directory is inclusive, which means that if a cache line is present in any cache, there must be an entry in this directory. When the directory is full or a conflict occurs, any previous entry's data is removed from all the caches and data entries are invalidated.

TABLE 1. SUMMARY OF MULTI-CMP SYSTEM CONFIGURATION.

| | | | |
|---|---|---|---|
| Cores. | | Number | 4 |
| | | Architecture | x86-64, Pipelined |
| | | Frequency | 3.5Ghz |
| | | Block size | 64 Bytes |
| Private Caches | L1 | Assoc. | 4-way |
| | | Size | 32 KB (Shared I/D) |
| | | Access time | 1 cycles |
| | L2 | Assoc. | 4-way |
| | | Size | 128 KB |
| | | Access time | 3 cycles |
| | | Type | Exclusive with L1 |
| Shared L3 | | Assoc. | 8 |
| | | Size | 4MB, 4 Slices |
| | | NUCA Mapping | Static, interleaved by LSB |
| | | Bank Access Time | 5 cycles |
| Memory | | Capacity | 4GB |
| | | Access time | 300 cycles |
| | | Memory Controllers / BW | 1 per CMP/32GBs |
| Network | | Link Latency / Link Width | 1 cycle, 16B 4 virtual channels |
| | | On-Chip Topology | 2x2 mesh |
| | | Off-Chip | 1+1 chips / 2x2 chips, 16B |

TABLE 2. WORKLOADS CHARACTERISTICS
(AND ACRONYMS USED IN FOLLOWING FIGURES)

| | | |
|---|---|---|
| SPEC CPU2006 | Gamess (GA) | Integer[45] |
| | Gcc (GC) | |
| | Namd(NA) | Floating point [45] |
| | Sjeng (SJ) | |
| | Sphinx3(SH3) | |
| NPB | Gradient conjugate (CG) | Class B [46] |
| | Fast Fourier Transform (FT) | |
| | Integer Sort (IS) | |
| | LU diagonalization (LU) | |
| | multi-grid (MG) | |
| | Sp (SP) | |
| | Ua (UA) | |
| Parsec | Canneal (CA) | Simlarge [47] |
| | Fluidanimate (FL) | |
| | Streamcluster (ST) | |
| | x264 (X4) | |

The AMD *Hyper-Transport* like protocol currently available in the gem5 simulator has been heavily modified to meet our system configuration (namely, add support for CMP and incorporating a third level of cache. See [24] for more details).

A large modification was required to provide multiple core chips (CMP) and to mimic current complex cache hierarchies present in commercial systems [5][48][49]. In our study, we will refer to this protocol as *HTA*. For further details about this implementation and state transition tables of the coherence controllers, please refer to [24].

As indicated previously, there is a plethora of related work in the literature. Nevertheless, most of these ideas are focused on single chip systems. The extension to the multi-CMP case might require substantial changes in the original ideas. In *HTA* protocol, the fundamental premise is that it is designed for constrained-cost multi-CMP servers (and it is being used on a successful commercial product). In the literature there are some coherence protocols, such as [7], also focused on multi-CMP systems. Nevertheless, multiple timeout dependent core protocol actions (such as persistent request) and semi-centralized arbitration renders this protocol unstable for memory intensive applications [41].

Rainbow is conceived as an extension to [15] which was proposed assuming a single CMP coherence protocol, like many others [29]. A performance comparison with this kind of protocols seems not useful for the discussion (in the context of the class of system considered). Finally, self-invalidation protocols, such as [17], are not considered. The complexity of the software stack (and the programmer productivity) for general purpose systems is hard to reconcile with the software ideas. Even when this can be done, it is unfair to perform a comparison with a strictly hardware-based approach such as *Rainbow* (that requires no changes in software or memory consistency model). Note that gem5 does not perform "timing-first" simulation (will perform not only unsupported features of the hardware via functional simulation, but will also mask issues in the coherence protocol), so a complete redesign of the platform, including processor ISA, might be required to support them.

## 5. PERFORMANCE RESULTS

### 5.1 Dual-CMP system

To present the benefits of *Rainbow*, we will sweep the parameters of the coherence structures used. These parameters are related to the D|F-LLC and D|F-MEM in *Rainbow* and the probe filter directory in *HTA*. In any comparison the total storage capacity of the required structures in the whole server will be the same in both protocols.

Under the non-cost-effective assumption of having enough capacity to store all the coherence information, the performance of both protocols will be very similar. Both will have all the necessary block information in their structures and, when needed, this information will be accessed in order to deal with consequent requests. Figure 8 shows the execution time normalized to *HTA*, in a dual-CMP system using a probe filter directory with a capacity of 131k entries of coherence information per CMP, i.e. ~8.2MB per chip (>x2 LLC size). For the system characteristics, this value doubles the number of blocks that can be stored in the caches of each chip. Consequently, the directory conflict misses are negligible [19]. *Rainbow* protocol structures use the same area, but distributed between the D|F-LLC and the D|F-MEM. This means that the

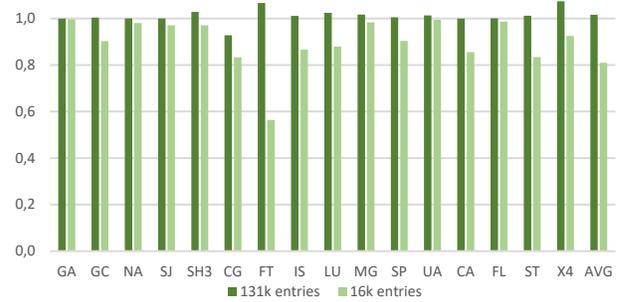

Figure 8. HTA-normalized execution time for *Rainbow* in a 2-CMP system, when there is enough space to track all coherence information (131k entries in probe filter) and when there is 1MB per CMP to track coherence information (16k entries in probe filter).

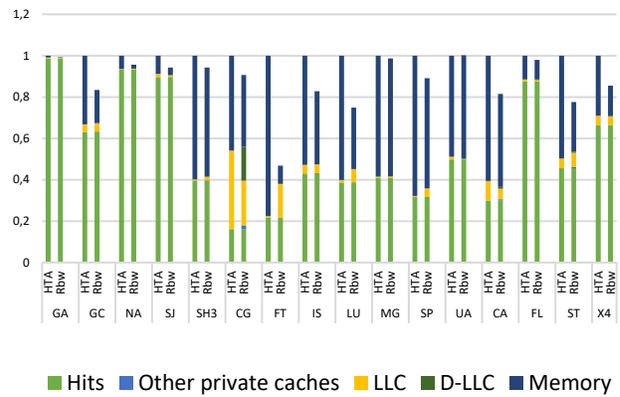

Figure 9. *HTA*-normalized average memory access latency for *Rainbow* in a 2-CMP system for 16k entries.

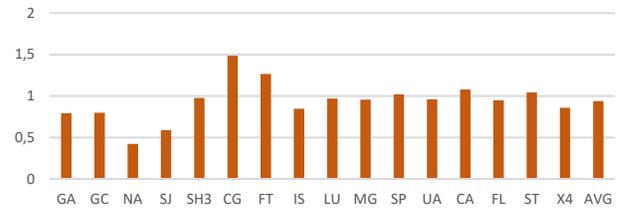

Figure 10. *HTA*-normalized link utilization of *Rainbow* protocols when using 1MB of coherence structures. (16K entries in probe filter) in a 2-CMP system.

D-LLC has capacity to track 8k entries each (there is one per L3 bank, i.e. four inside the CMP) and the D-MEM tracks 32k entries. This totals 64k entries which correspond to half of the space of HTA. The other half is space dedicated to F-LLC and F-MEM. As expected, both protocols show similar behavior (Figure 8). Since there are no external invalidations due to directory misses, *HTA* performance is optimal. *Rainbow* in most cases achieves almost the same performance. In some cases, it is slightly lower (~5%) due to the reconstruction process in private-to-share block promotion (which is unnecessary in *HTA*).

Nevertheless, when coherence resources are more realistically set, the behaviour of both protocols will differ. In this particular case we limit them to 1MB (i.e. ¼ LLC size, which is maintained constant across all the results). In *HTA*, for the system configuration used, each probe filter directory can track 16k blocks. In *Rainbow,* first we tune F-LLC and F-MEM sizes to limit the probability of false positives to below 5% (assuming a random access pattern). Then, to equalize the cost, D-LLC will have capacity to track 512 blocks while the D-MEM will track 4k blocks. Then, *Rainbow* directories in the CMP will track 6k blocks: 2k in the four LLC controllers and 4k in the memory controller. The remaining storage is used by F-LLC and F-MEM.

Whereas *HTA* is degraded due to the external invalidations from the probe filter evictions, *Rainbow* barely notices the effect. In some multithread applications with large sharing degree, such as FT, the performance loss of *HTA* is nearly 50%. In any case, for most of the applications, the number of misses in the on-chip caches is increased in the *HTA* protocol due to the invalidations caused by the probe filter directory evictions. As we can see, even non-multithreaded applications, such as SPEC-based ones, show a minor performance degradation due to the Linux NUMA policy mapping.

To support the discussion of the effects of these external invalidations, Figure 9 presents the *HTA*-normalized components of the average memory latency for 16k entries. The contribution of hits in the private caches is similar in both protocols. However, the external invalidations in *HTA* induce many extra memory accesses, consequently the contribution of the memory is much greater than in *Rainbow*. In most cases, when reducing the size of the coherence structures, the number of conflicts or capacity misses due to lack of space increases. In *Rainbow* protocol the absence of invalidations induced by directory conflicts makes it immune to such effects. The higher number of reconstructions is barely noticeable in the access latency.

The most evident *Rainbow* drawback is the increment in the system traffic due to the D-LLC and/or D-MEM entry reconstructions. Since contention effects due to this are already included in the previous results, it is interesting to compare the bandwidth utilization, as a proxy of Energy requirements, Figure 10 provides the average bandwidth utilization across the whole system. On average, the traffic generated by *HTA* is greater than that of *Rainbow*. The effect of the invalidations caused by the lack of space in the directory is larger than the effect of the higher number of reconstructions needed by *Rainbow*. However, it is true that in applications where *Rainbow* achieves better performance, it has greater bandwidth requirements. Note that for the class of system considered (general purpose servers), other metrics such as Energy Delay Product (EDP) or Energy Delay Square Product (ED2P) are more useful to evaluate the system efficiency. Under such metrics, *Rainbow* is more efficient in all applications, having, on average, an EDP ~25% and ED2P 40% smaller than *HTA*.

## 5.2 Protocol Behavior under highly Stressed configuration

Next, we will reduce coherence storage availability to unrealistically low values. In this way we can infer how each protocol reacts under demanding scenarios that are not necessarily covered by the workloads tractable via simulation. We will divide by four the previously considered sizes. This means that the probe filter will be able to track 4k entries, while *Rainbow* will dedicate half of that space to its filters and it will maintain a directory in the D-LLC of 128 entries. The D-MEM will hold 1k entries to track the blocks inside the CMPs of the system.

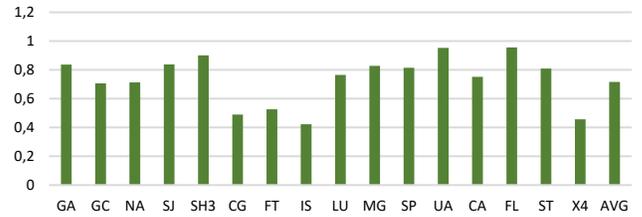

Figure 11. *HTA*-normalized execution time for *Rainbow,* when there is only 256KB per CMP to track coherency information (4k entries in probe filter) in a 2-CMP system.

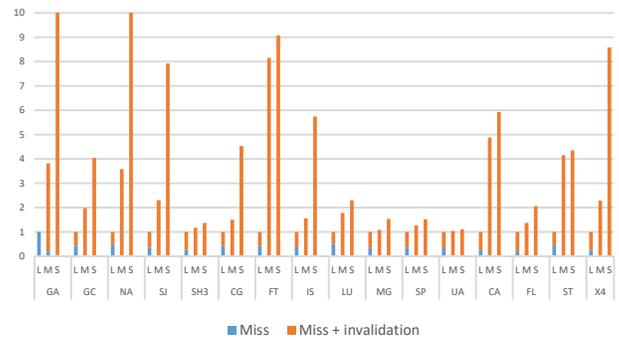

Figure 12. Normalized number of misses in *HTA* when varying the probe filter directory (L, large: 131k; M, medium: 16k; S, small: 4k) in a 2-CMP system.

Under such an extreme configuration, as Figure 11 shows, Rainbow reacts much better than *HTA*. The gap between the two protocols is now even more noticeable across all benchmarks. In multithreaded applications, the performance advantage of *Rainbow* is much greater than before, with two applications above 60% of execution time reduction. Even non-multithreaded workloads are critically affected by the increment in external invalidations.

As we reduce the amount of storage dedicated to coherence information, we see how the *HTA* protocol is negatively affected due to the increase in the number of conflicts occurring in its directory. Figure 12 shows how, in *HTA*, the number of on-chip misses grows as the probe filter size is reduced.

In contrast, *Rainbow* reacts significantly better to this resource scarcity. Small D-LLC and D-MEM also means more misses in the two structures and consequently the number of reconstructions will be increased. Since there is substantial diversity in reconstruction frequency for different benchmarks, we choose to compare bandwidth utilization as an indirect metric. Figure 13 presents the results for the three configurations normalized to the realistic one. For reference, we include that metric also for *HTA.* For *Rainbow,* in most cases the traffic increment is barely noticeable. This is an indicator that both F-LLC and F-MEM successfully prevent unnecessary snoops. Focusing on the details, it is important to highlight the complementary nature of both structures when filtering traffic. When there is a miss in the D-LLC and a hit in F-LLC, the reconstruction multicast is constrained to the chip. When a request reaches the memory controller and there is a miss in D-MEM and a hit in F-MEM, only the corresponding LLC controller in each chip will be snooped, which in most cases will be filtered out by a miss in F-LLC. The worst-case scenario (i.e. snooping all the private caches) is infrequent, even in the most extreme configurations. For the applications under study, D-MEM is less used than F-MEM. While this opens the opportunity to adjust their capacities, we left them fixed (adjusting filter capacity to 5% false positives). Although it seems unfair to perform such an optimization, there is room for improvement. In general, depending on system workload, it is feasible to split the capacity of directory and filter dynamically, both in the memory controller and LLC controller. Notwithstanding, this possibility has not been explored in this work.

### 5.3 Scalability to larger CMP count.

Figure 14 shows how memory latency behaves on a 4-chip server using the same large, medium and small directory configurations as in the previous subsection (other parameters of the systems are unaltered). Similar behavior to the 2-CMP system can be observed. While *HTA* is negatively affected by the reduction in the directory size, *Rainbow* hides this from the perspective of memory access time. The contribution of each type of coherence agents to the average latency stays steady across different capacities of D|F-LLC and D|F-MEM, meaning that they are overprovisioned for most applications.

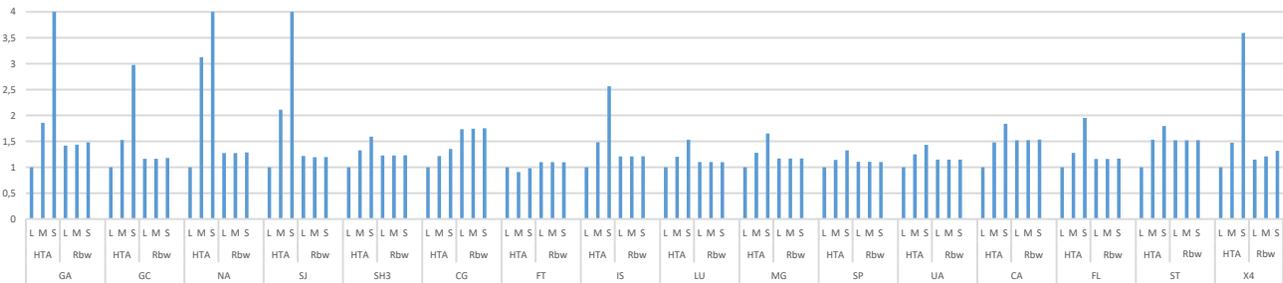

Figure 13. Large-size normalized Bandwidth utilization of the two protocols while varying the probe filter directory (large: 131k; medium: 16k; small: 4k) in a 2-CMP system.

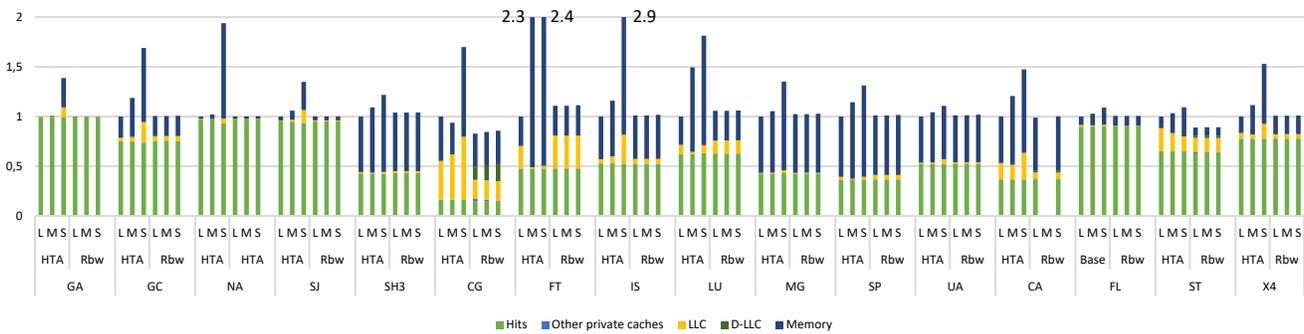

Figure 14. Large-size normalized memory latency in a 4-CMP system.(large: 131k; medium: 16k; small: 4k) and LLC (full:4MB per CMP, half: 2MB per CMP) in a 4-CMP system.

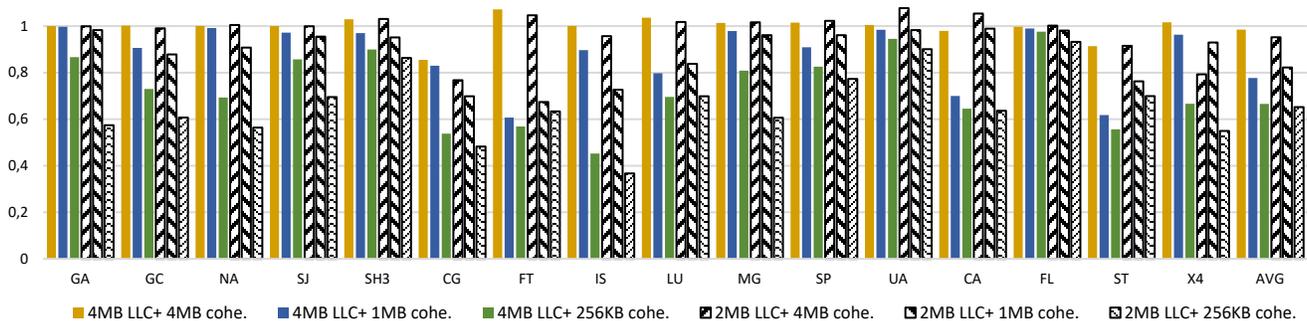

Figure 15. HTA-normalized execution time while varying the probe filter directory (large: 131k; medium: 16k; small: 4k) and LLC size (full:4MB per CMP, half: 2MB per CMP) in a 4-CMP system.

It is important to notice that when duplicating the number of chips, we are also duplicating the aggregate number of cache blocks to be tracked by the *HTA* directory or the D-MEM structures, but the number of memory controllers are also duplicated so the total storage available is doubled too.

Duplicating the number of cores also means dividing the workload across more cores. Therefore, to exert more pressure on the memory controller coherence structures, we provide performance results with full LLC size (i.e. 4MB per CMP) and reduced LLC size (i.e. 2MB per CMP). Both configurations have been tested with the three different capacities for the coherence structures used previously.

Figure 15 shows how *Rainbow* is slightly affected when using the large directory because of the higher pressure on the D|F-LLC and D|F-MEM due the increment in the number of data blocks that must be tracked. This causes a higher number of false positives in F-LLC and F-MEM. This slightly increases memory latency even in the unrealistic case. However, as we reduce the coherence structures, the *HTA* protocol is critically affected by external invalidations in private caches. On average, *Rainbow* is ~25% and ~65% faster than *HTA* with regular or constrained coherency structures respectively, both for 2MB and 4MB LLC caches.

## 6. CONCLUSIONS

We have proposed a composable coherence protocol for multi-CMP systems. The use of colored token counting both simplifies the maintenance of the invariant SWMR and allows the isolation of complexity from performance or power optimizations. The approach facilitates the composability according the different coherence domains that appears in this type of systems. The experimental results suggest that it is an efficient mechanism to tackle the system complexity. Such goal has been achieved through the introduction of a structure composed of both a non-exact directory and a filter associated with each of the different domains. Thus, associating one of these structures with the last level of cache (D|F-LLC) reduces the memory and communication requirements within the chip and through another structure associated with the memory controller (D|F-MEM), it does the same at the system level. The results obtained from applications running under full system simulation indicates that this proposal can be useful to increase the core count of multi-chip servers keeping attained production costs. Current market evolution (and performance cost figures) suggest that large-count multi-CMP might gain traction in near-future computing infrastructures, especially with trends in current workloads memory requirements and DRAM cost.

**Lucia G. Menezo** received her B.S. and M.S. from the University of Basque Country in 2007. In 2014, she received her Ph.D. degree from the University of Cantabria, where she works as a researcher since then. Her research interests focus on the memory hierarchy, mainly on cache coherence protocols for chips multiprocessor (CMPs).

**Valentin Puente** received the B.S., M.S. and Ph.D. degrees from the University of Cantabria, Spain, in 1995 and 2000, respectively. He is currently an associate professor of computer architecture at the Department of Computers and Electronics of the same University. His research interests focus on memory hierarchy design and the impact that incoming technology changes might have on it.

**Jose A. Gregorio** received the B.S., M.S. and Ph.D. degrees in physics (electronics) from the University of Cantabria, Spain, in 1978 and 1983, respectively. He is currently a professor of computer architecture in the Department of Computers and Electronics in the same university. His main research interests focus on chip multiprocessors (CMPs) with special emphasis on the memory subsystem, interconnection network and coherence protocol of these systems.